\documentstyle[psfig,sf99proc]{article}

\title{HST study of the Stellar Populations within 30~pc of SN~1987A}

\authors{
Martino Romaniello\affilmark{1}, Nino Panagia\affilmark{2} and
Salvatore Scuderi\affilmark{3}
}

\affiltexts{
\affiltext{1}{ESO, Karl-Schwarzschild-Strasse~2, D-85748 Garching bei
M\"unchen, Germany.}
\affiltext{2}{STScI, 3700 San Martin Drive, Baltimore, MD~21218. Affiliated to
the Astrophysics Division, Space Science Department of ESA.}
\affiltext{3}{Osservatorio di Catania, Viale A. Doria 6, Catania, Italy.}
}

\firstauthor
{MR}
{mromanie@eso.org}

\authorsADS{%
Romaniello, M.;
Panagia, N.;
Scuderi, S.
}

\affiliationsADS{%
AA(European Southern Observatory)
AB(Space Telescope Science Institute, affiliated to the ESA Astrophysics
Division)
AC(Osservatorio di Catania)
}

\begin{document}

%
\begin{abstract}
We have studied the stellar populations in a region of 30~pc around
SN~1987A in the Large Magellanic Cloud using multi-band \emph{HST}-WFPC2
images. 
The effective temperature, luminosity and reddening of each detected star were 
determined by fitting the measured broad band magnitudes to the ones
calculated with model atmospheres. The resulting HR diagram reveals the
presence of stars with ages between 1 and 150~Myrs, superposed on a much older
field population. The youngest stars in the field appear to be T~Tauri stars,
characterized by strong H$\alpha$ excesses. 
The Star Formation Rate has increased monotonically in the last 8~Gyrs and 
the star formation activity is still very high at present.
\emph{Young} stars of low and high mass have different spatial distributions.
The latter ones are strongly concentrated in the vicinities of the Supernova,
whereas the previous ones are more evenly distributed on the field. Hence, the
Mass Function varies on a scale of a few parsecs. However, averaging over the
entire area, one can define an Initial Mass Function, well fitted by a
power-law with a slope $\alpha\la-2.55$. The uncertainty on the IMF due to
incompleteness in identifying T~Tauri stars is discussed. 
\end{abstract}

\section{Introduction}
SN~1987A is located at the SW edge of the Tarantula Nebula, some
$20^\prime$ away from its center. The whole area contains a large
number of early type stars interspersed with HII regions and SNR shells.
The OB association closest to SN~1987A is LH~90, which is located about
$5^\prime$ to the NE of  the supernova (Lucke \& Hodge 1970) and whose
age is much younger than that of SN~1987A progenitor, {\it i.e.}, about
4~Myrs as compared with the 10-11~Myrs as estimated for Sk~-69~202
({\it e.g.} Van Dyk, Hamuy and Mateo, 1998). It is clear that the study of
SN~1987A neighborhood offers a unique opportunity to place the supernova
explosion in the proper context of stellar evolution and the evolution of
stellar populations.

\section{Observations and Data Reduction}
SN~1987A has been observed with various instruments on board HST since
1990. In particular, as part of the SINS project ({\bf S}upernova {\bf
IN}tensive {\bf S}tudy, PI Kirshner), we have now a series of
multifilter images that give an excellent coverage over an area of
about 130$^{\prime\prime}$ radius, \emph{i.e.} about 30~pc, centered
on SN~1987A. 
Here we present the results of the analysis of the SINS images taken
with the WFPC2 camera, using the F255W, F336W, F439W, F502N, F555W,
F656N, F675W, F814W filters (in the following we shall refer to the broad
bands as UV, U, B, V, R and I although they do not coincide with any of
the canonical ground based color systems). We also use an archival F656N
image taken in early 1994 under project \#5203 (PI Trauger).

After processing the observations through the PODPS (Post Observation
Data Processing System) pipeline for bias removal and flat fielding,
we have removed cosmic ray hits by combining the two available images in each
filter. Finally, we performed aperture photometry following the
prescriptions of Gilmozzi (1990) with the
refinements as described by Romaniello (1998). The flux
calibration was obtained using the internal calibration of the WFPC2,
which is typically accurate to better than $\pm$ 5\%.
In this way we obtained the photometry of a total of 21,955 stars. More than
15,000 of them have a photometric accuracy better than 0.1~{\it mag} in the
V, R and I filters. This number drops to 6,825 in the B band and only
786 stars have a UV filter uncertainty smaller than 0.2~{\it mag}.

\section{The HR Diagram}
\begin{figure}[!ht]
\centerline{\psfig{file=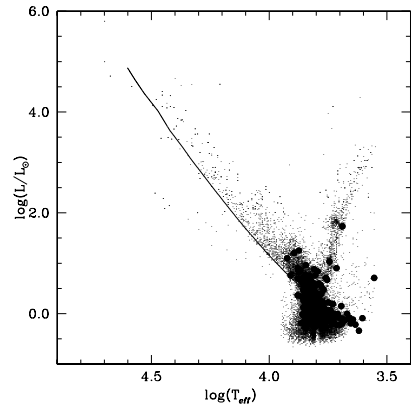,width=0.7\linewidth}}
\caption{HR diagram for the 21,955 stars in the field around SN~1987A. T~Tauri
stars identified through their H$\alpha$ excess are shown as dots. The full
line shows the location of the ZAMS for $Z=Z_\odot/3$ from the models of
Brocato and Castellani (1993) and Cassisi et al (1994).}
\label{fig:hr}
\end{figure}

The large number of bands available (6 broad band filters) which cover a
wide baseline (more than a factor of 3 in wavelength, extending from
2550\AA\ to 8140\AA) provide us with a sort of {\it wide-band
spectroscopy} which defines the continuous spectral emission
distribution of each star quite well.  Therefore, by comparison with
model atmospheres (Bessel et al, 1998), one can fit the 6~band
observations of each stars and solve for 3 unknowns simultaneously,
namely the effective temperature, $T_{eff}$, the reddening, $E(B-V)$,
and the angular radius, $R/D$.  In practice, this can be done only for
stars with effective temperatures higher than about 10,000~$K$ and between
8,500 and 6,500~$K$ (for details, see Romaniello, 1998). Therefore, we first
solved for the full set of parameters, $T_{eff}$, $E(B-V)$, and $R/D$ only
for stars suitably selected on the basis of reddening-free colors.  For each of
the remaining stars, we adopt the average reddening of its first
neighbors and solve for only two parameters,  $T_{eff}$, and $R/D$.
Finally the stellar luminosity is computed from the derived $T_{eff}$
and $R/D$, adopting a distance to SN~1987A of 51.4~$kpc$ (Panagia 1998).
The resulting $\log\left(L/L_\odot\right)$~vs.~$\log\left(T_{eff}
\right)$ plot (HR diagram) is shown in Figure~\ref{fig:hr}. 

By comparing the R band magnitudes with the ones measured in the 
H$\alpha$ narrow band filter we identified the stars with strong 
H$\alpha$ excess ($R-m(H\alpha)>0.25$, i.e. $W_{eq}(H\alpha)>8$\AA).
We identify the luminous and bright ones (5 stars), which are near the MS,
as Be stars whereas we believe that the redder and fainter ones (488 stars)
are T~Tauri stars, i.e. Pre-MS stars with circumstellar material,
remnant of their proto-stellar cocoons. The vast majority of these stars also
show a U-band excess, \emph{i.e.} very blue U$-$B colour given their B$-$I
one. Again, this is interpreted as due to the accretion of material from the
disk. With this criterion, we identify 850 T~Tauri candidates.

An inspection of the HR diagram confirms the early findings of Walker \&
Suntzeff (1990) and Walborn et al (1993) and reveals that:

\begin{itemize}
  \item The distribution of stars in the HR diagram is clearly 
bound toward high temperatures, identifying a  ZAMS that corresponds to 
a metallicity $Z\simeq Z_\odot/3$.
  \item The positions of the most luminous blue stars fall on
isochrones corresponding to ages around 10-12~Myrs, which make them
coeval to SN~1987A progenitor and Star~2 (Scuderi et al 1996).
  \item There are a number of stars at intermediate luminosities and
temperatures (say, $\log(L/L_\odot\sim 2-4$ and $\log(T_{eff}\sim3.8-4.2$)
that indicate distinct stellar generations, with ages in the range 
40-150~Myrs.
  \item The lower MS and the red giants are mostly old
populations, consistent with a metallicity either identical to, or
slightly lower than the one of the young components.  No single age
can explain the distribution of the old population, and stellar
generations between 600~Myrs and 6~Gyrs are required to account
for the observations.
  \item Among  the stars with substantial H$\alpha$ excess we identify 488
strong T~Tauri stars, whose positions in the HR diagram indicate ages from
1-2~Myrs up to 10-20~Myrs, according the Pre-MS isochrones of Siess et
al (1997).
\end{itemize}

\section{The spatial distribution}
\begin{figure}[!ht]
\centerline{\psfig{file=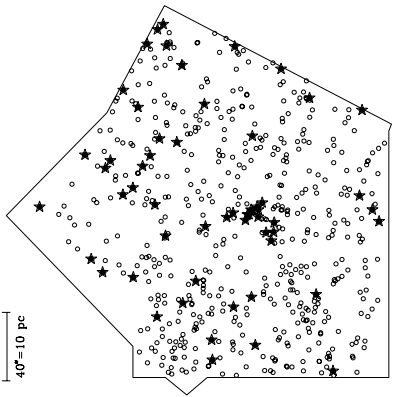,width=0.7\linewidth}}
\caption{Spatial distribution of massive (star symbol) and Pre-Main Sequence
(circles) stars.}
\label{fig:sd}
\end{figure}

As shown in  Figure~\ref{fig:sd}, the spatial distribution of early type,
massive stars and the one of PMS stars does not correlate well with each
other. The ratio of low mass to high mass stars belonging to the most recent
episode of star formation shows big variations across the field. For example,
it is of the order of unity in the central area, where Supernova~1987A is,
while elsewhere it varies between roughly 7 (south-east of SN~1987A) and 28
(south-west of SN~1987).

In general, there is a lack of massive stars in the south-west corner, where
the T~Tauri stars are most most numerous.These differences are highly
significant because they greatly exceed the simple Poissonian fluctuations. An
experimental check of this fact is provided by similar statistics on the
number of stars belonging to the Red Giant clump. They are part of a much
older population which, therefore, should be uniformly distributed over the
field.  Indeed, the observed number densities for these stars show
fluctuations perfectly in agreement with Poisson statistics. 

\emph{The almost anti-correlation of spatial distributions of high and low mass
stars of a coeval generation indicates that star formation processes for
different ranges of stellar masses are rather different and/or require
different initial conditions}. An important corollary of this result is that
the very concept of an Initial Mass Function seems not to have validity in
detail, but may rather be the result of a random process, so that it could
make sense to talk about an average IMF over a suitably large area
in which all different star formation processes are concurrently operating.

\section{Star Formation History and Initial Mass Function}
\begin{figure}[!ht]
\centerline{\psfig{file=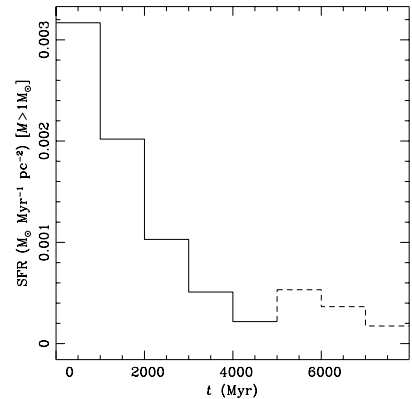,width=0.7\linewidth}}
\caption{SFR for the field around SN1987A. Look back times
greater than 5~Gyr are affected by incompleteness (dashed histogram).}
\label{fig:sfh}
\end{figure}

As we have seen, the observations cannot be explained in terms of a single
generation of stars. We have recovered the Star Formation History of the field
by assigning an age (and mass) to every observed star by comparing its
position on the HR diagram to the evolutionary models by Brocato and Castellani
(1993) and Cassisi, Castellani and Straniero (1994). The result is shown
in Figure~\ref{fig:sfh}

An inspection of Figure~\ref{fig:sfh} shows that the Star
Formation Rate has been increasing in the past 5~Gyr by a factor of 15 if
the value of the bin centered at 4.5~Gyr is taken as representative of the
star formation activity in the remote past. However, this bin too may be
affected by incompleteness problems. In a more conservative way, we can take
the SFR value at 3.5~Gyr as representative, in which case the enhancement is
a factor of 6.

\begin{figure}[!ht]
\centerline{\psfig{file=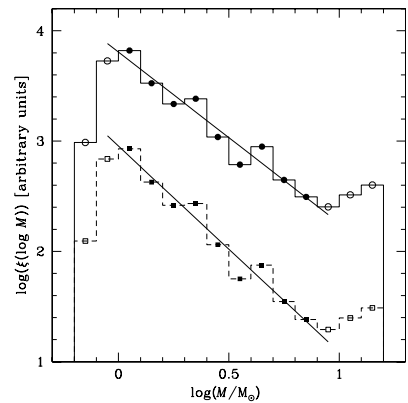,width=0.7\linewidth}}
\caption{IMF including only stars with H$\alpha$ excess (full line)
or also those with U-band excess (dashed line).}
\label{fig:imf}
\end{figure}

The derived IMF is shown in Figure~\ref{fig:imf}. Let us stress here that this
is a \emph{spatial and temporal average} resulting from the superposition
of different generation of stars and needs not to be valid for any of them
individually.

A least-square fit in the $1\leq M/\mathrm{M}_\odot\leq 7$ interval, where the
sample is not affected by incompleteness, yields a slope of $\alpha=-2.55$ if
only the stars with H$\alpha$ excesses are included in the young population
and a value of $\alpha=-2.87$ if also the stars with U-band excesses are added.

Full account of this work can be found in Romaniello (1998).

\section{References}
\reference Bessel, M.S., Castelli, F., and Plez, B. 1998, AA, 333, 231.
\reference Brocato, E., \& Castellani, V., 1993, ApJ, 410, 99.
\reference Cassisi, S., Castellani, V., \& Straniero, O., 1994, A\&A, 
	282, 753.
\reference Gilmozzi, R., 1990, ``Core aperture photometry with the WFPC'', 
      	STScI Instrument Report WFPC-90-96.
\reference Lucke, P.B., \& Hodge, P.W., 1970, AJ, 75, 171.
\reference Panagia, N., 1998, Invited Talk at the Workshop {\it Views on 
  Distance Indicators}, ed. F. Caputo, Mem.S.A.It., in press.
\reference Romaniello, M. 1998, PhD Thesis, Scuola Normale Superiore, Pisa.
\reference Scuderi, S., Panagia, N., Gilmozzi, R., Challis, P.M. \& Kirshner,
        R.P., 1996, ApJ, 465, 956. 
\reference Siess, L., Forestini, M., \& Dougados, C. 1997, A\&A, 325, 556.
\reference Van Dyk, S., Hamuy, M., \& Mateo, M., 1998, in {\it SN~1987A: 
  Ten Years Later}, eds. M.M. Phillips and N.B. Suntzeff, ASP Conf. Ser., 
  in press.
\reference Walborn, N.R., Phillips, M.M., Walker, A.R., \& Elias, J.H., 
	1993, PASP, 105, 1240. 
\reference Walker, A.R., \& Suntzeff, N.B., 1990, PASP, 102, 131.

\vspace{-1mm} 

\end{document}